\documentclass[aps,prl,twocolumn,superscriptaddress,shopacs]{revtex4}
\usepackage{graphicx}
\usepackage{bm}
\usepackage{gensymb}
\usepackage[none]{hyphenat}
\usepackage{epstopdf}
\usepackage{multirow}
\usepackage{float}
\usepackage{tabularx}
\usepackage{gensymb}
\usepackage{color}
\usepackage{siunitx}
\usepackage{xr}
\usepackage{amsmath}
\begin{document}

\title{Incommensurate magnetism near quantum criticality in CeNiAsO}
\author{Shan Wu}
\email{shanwu@berkeley.edu}
\affiliation{Department of Physics and Astronomy and Institute for Quantum Matter, Johns Hopkins University, Baltimore, Maryland 21218, USA}
\affiliation{Department of Physics, University of California Berkeley, Berkeley, California 94720, USA}

\author{W. A. Phelan }
\affiliation{Department of Physics and Astronomy and Institute for Quantum Matter, Johns Hopkins University, Baltimore, Maryland 21218, USA}

\author{L. Liu}
\affiliation{Department of Physics, Columbia University, New York, New York 10027, USA}

\author{J. R. Morey}
\author{J. A. Tutmaher}
\affiliation{Department of Physics and Astronomy and Institute for Quantum Matter, Johns Hopkins University, Baltimore, Maryland 21218, USA}

\author{J. C. Neuefeind}
\affiliation{Oak Ridge National Laboratory, Chemical and Engineering Materials Division, Oak Ridge, Tennessee 37831, USA }

\author{Ashfia Huq}
\affiliation{Oak Ridge National Laboratory, Neutron Scattering Division, Oak Ridge, Tennessee 37831, USA }

\author{Matthew B. Stone}
\affiliation{Oak Ridge National Laboratory, Neutron Scattering Division, Oak Ridge, Tennessee 37831, USA }

\author{M. Feygenson }
\affiliation{Juelich Centre for Neutron Science, Forschungszentrum Juelich GmbH, 52425 Juelich, Germany
}

\author{David W. Tam}
\affiliation{Department of Physics and Astronomy, Rice University, Houston, Texas 77005, USA}

\author{Benjamin A. Frandsen}
\affiliation{Department of Physics and Astronomy, Brigham Young University, Provo, Utah 84602, USA}

\author{Benjamin Trump}
\affiliation{NIST Center for Neutron Research, National Institute of Standards and Technology, Gaithersburg, Maryland 20899, USA}

\author{Cheng Wan}
\affiliation{Department of Physics and Astronomy and Institute for Quantum Matter, Johns Hopkins University, Baltimore, Maryland 21218, USA}

\author{S. R. Dunsiger}
\affiliation{Department of Physics, Simon Fraser University, Burnaby, B.C., Canada}

\author{T. M. McQueen}
\affiliation{Department of Physics and Astronomy and Institute for Quantum Matter, Johns Hopkins University, Baltimore, Maryland 21218, USA}
\affiliation{Department of Materials Science and Engineering, Johns Hopkins University, Baltimore, Maryland 21218, USA}

\author{Y. J. Uemura}
\affiliation{Department of Physics, Columbia University, New York, New York 10027, USA}

\author{C. L. Broholm}
\affiliation{Department of Physics and Astronomy and Institute for Quantum Matter, Johns Hopkins University, Baltimore, Maryland 21218, USA}
\affiliation{Department of Materials Science and Engineering, Johns Hopkins University, Baltimore, Maryland 21218, USA}
\affiliation{Oak Ridge National Laboratory, Quantum Condensed Matter Division, Oak Ridge, Tennessee 37831, USA }

\date{\today}
\begin{abstract}
We report the discovery of incommensurate magnetism near quantum criticality in CeNiAsO through neutron scattering and zero field muon spin rotation.  For $T <T_{N1}$ = 8.7(3) K, a  second order 
phase transition yields an incommensurate spin density wave with wave vector  $\textbf{k} = (0.44(4), 0, 0)$. For $T < T_{N2}$ = 7.6(3) K, we find co-planar commensurate order with a moment of $0.37(5)~\mu_B$, reduced to 30 \% of the saturation moment of the $|\pm\frac{1}{2}\rangle$ Kramers doublet ground state, which we establish through inelastic neutron scattering. 
Muon spin rotation in $\rm CeNiAs_{1-x}P_xO$ shows the commensurate order only exists for x $\le$ 0.1 so  we infer the  transition at $x_c$ = 0.4(1) is between an incommensurate longitudinal spin density wave and a paramagnetic Fermi liquid.

\end{abstract}

\maketitle

The competing effects of intra-site Kondo screening and inter-site Ruderman-Kittel-Kasuya-Yosida (RKKY) interactions in rare earth intermetallics epitomize the strongly correlated electron problem. While the N\'{e}el and Kondo lattice limits are well 
understood \cite{Doniach}, the transition between them is far from. It involves an increase in the volume enclosed by the Fermi surface (FS) as the $4f$ electron is incorporated on the Kondo lattice side of the transition\cite{Stewart, Coleman}. Deviations from the $\rho\propto T^2$ dependence of resistivity is  interpreted as indicative of the associated quantum criticality, which is denoted as ``local" because it involves the entire FS. In support of this concept, compounds with the requisite transport anomalies have  been discovered where  physical properties that involve averages over distinct regions of momentum space have related critical exponents. The eventual transition to magnetic order when RKKY interactions dominate can coincide with the localization transition or occur within  the large or small FS phases. Clearly the nature of the corresponding quantum critical point is strongly affected as magnetic ordering is  momentum selective and breaks time reversal symmetry. 

Exploration of model systems is essential to uncover the overall phase diagram of this complex strongly correlated regime. CeCu$_{6-x}$Au$_x$ provided a first example of local criticality. de Haas-van Alpen measurements provide  evidence for an abrupt rearrangement of the FS in CeRhIn$_5$ at  2.25 GPa \cite{Jiao20012015,Park2006HiddenMA, shishido2005}  . A step change in the Hall coefficient of YbRh$_2$Si$_2$ coupled with anomalous and yet unexplained critical exponents at the field driven ferromagnetic transition have been interpreted as evidence the magnetic and the electron localization transitions coincide\cite{Collin2012, friedemann2009detaching,PMID,Paschen_nature2004,Friedemann_2010}. Each compound adds unique insights and distinct experimental opportunities.

Isostructural to the 1111 iron pnictides, CeNiAsO is an exciting new addition to the landscape of strongly correlated electron systems \cite{Luo2012}. Magnetically ordered at low$-T$ and ambient pressure,  substitution of P for As or  pressure drives CeNiAs$_{1-x}$P$_x$O to a  paramagnetic Fermi-liquid. Non-Fermi-liquid transport is found up to the critical pressure $P_c$ = 6.5 kbar and the critical composition  $x_c=0.4(1)$ and a sign change in the Hall coefficient at $P_c$ indicates FS reconstruction \cite{Luo2014}. CeNiAsO differs from other systems studied to date in having two magnetic phase transitions \cite{Luo2012}. 

\begin{figure*}
\includegraphics[width=2.\columnwidth,clip,angle =0]{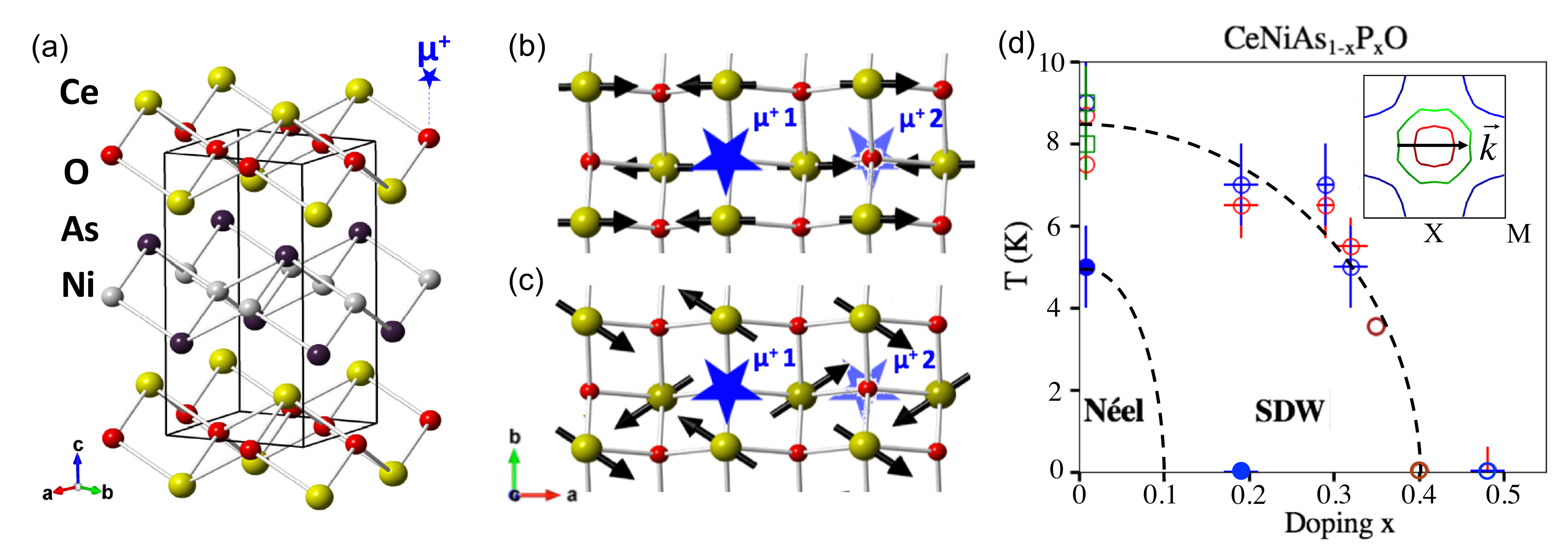}
\caption{\label{spin} (a) Crystallographic structure of CeNiAsO, and spin structure for  $T_{N2}<T < T_{N1}$ (b) and $T < T_{N2}$ (c).  Blue stars indicate the single crystallographic muon site. Two equivalent muon sites above and below oxygen site become inequivalent within the magnetically ordered state. (d) Temperature-doping phase diagram. Red, blue, and green symbols are from  specific heat, $\mu$SR, and neutron data respectively. Brown dots are from Luo et al. \cite{Luo2014}. We assign open (closed) symbols to the higher  (lower) $T$ transition. The inset to (d) shows the $q_z=0$ small Fermi surface excluding $4f$ electrons. The arrow shows the magnetic wave vector, which connects extended areas of the Fermi surface. The dashed lines are guides to the eye. }
\label{structurepd}
\end{figure*}

In this letter we determine the corresponding magnetic phases and examine their interplay with FS reconstruction. We show the upper transition is to an incommensurate  longitudinal SDW state with wave vector  $\textbf{k} = (0.44(4),0,0)$ that closely matches the umklapp wave vector ($2 {\bf k}_f$) of the small FS. The second transition yields co-planar commensurate order with  a low$-T$ ordered moment  reduced to 30\% of the saturation moment of the nominal $|\pm\frac{1}{2}\rangle$ Kramers doublet ground state. P doping suppresses the commensurate phase but retains the SDW  perhaps  all the way to the critical concentration where the FS grows to include $4f$ electrons.

We probed the magnetism of CeNiAsO through magnetic neutron scattering on the  NOMAD and POWGEN diffractometers \cite{Neuefeind,huq2011powgen} and  on the SEQUOIA \cite{sequoia} spectrometer  at the Spallation Neutron Source. 
For complementary real space information we used muon spin rotation ($\mu$SR) at the M15 beam line at TRIUMF. Specific heat measurements were conducted on a 14 Tesla Quantum Design PPMS with a dilution fridge insert. 

Fig. \ref{structurepd}(a) shows the tetragonal structure of CeNiAsO where magnetism is associated with $\rm Ce^{3+}$ sandwiching a square lattice of oxygen. The structure and the single phase nature of the sample was ascertained by Rietveld refinement of high resolution neutron diffraction data (see SI). The specific heat data in Fig. \ref{tdp}(d) show   shoulder-like  anomalies indicating two second order phase transitions  at $T_{N1}$ = 9.0(3) K and $T_{N2} $ = 7.6(3) K. The inferred critical temperatures are consistent with previously published specific heat data with sharper peaks indicating higher purity\cite{Luo2012}. The rounded maxima shift towards lower $T$ and approach each other in a field of $\mu_0 H = 14$ T as for two distinct antiferromagnetic phases.  

\begin{figure*}
\includegraphics[width=2.\columnwidth,clip,angle =0]{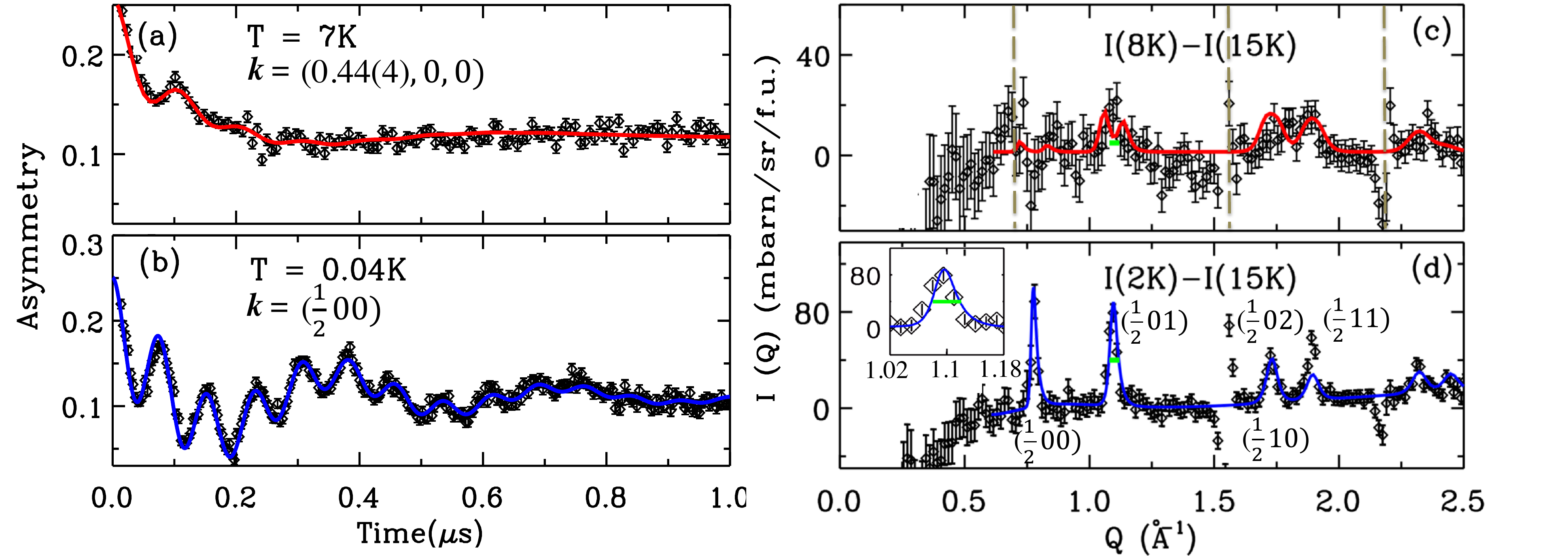}
\caption{\label{fig2} { (a-b) Zero-field longitudinal configuration $\mu$SR spectra at $T$ =  7 K and 0.05 K}. The colored  lines were calculated for the magnetic structures of Fig. \ref{spin}. (c-d) Diffraction patterns collected at $T$ = 2 K and 8 K on NOMAD, after subtracting $T$ = 15 K data as a measure of nuclear diffraction. Red and blue lines correspond to the spin configurations in Fig.~\ref{spin}. The grey dashed lines in (c) mark the nuclear Bragg positions, where thermal expansions gives rise to a peak-derivative anomaly. In (d) the horizontal green bar at $Q = 1.1 \AA^{-1}$ indicates the instrument resolution of 0.04 $ \rm \AA^{-1}$ as detailed in the inset.}
\end{figure*}

To determine their nature, we use zero field $\mu$SR  in the longitudinal configuration \cite{schenck,lee1998muon}. Fig. \ref{fig2} shows muon spin precession indicative of a well defined static internal field for  $T<T_{N1}$. A qualitative change in  the $\mu$SR profile for $T<T_{N2}$ indicates two distinct magnetic phases. For $T_{N2}< T <T_{N1}$, muons sample the broad spectrum of local fields generated by an incommensurate SDW \cite{tomo}. 
For $T < 6$ K,  the signal is oscillatory (Fig. \ref{fig2} (b)) with a  beating pattern that indicates two distinct precession frequencies and commensurate magnetism. These patterns can  be fitted by  magnetic structures that are consistent with the neutron data and a single crystallographic  muon stopping site.

\begin{figure}[t]
\includegraphics[width=.95\columnwidth,clip,angle =0]{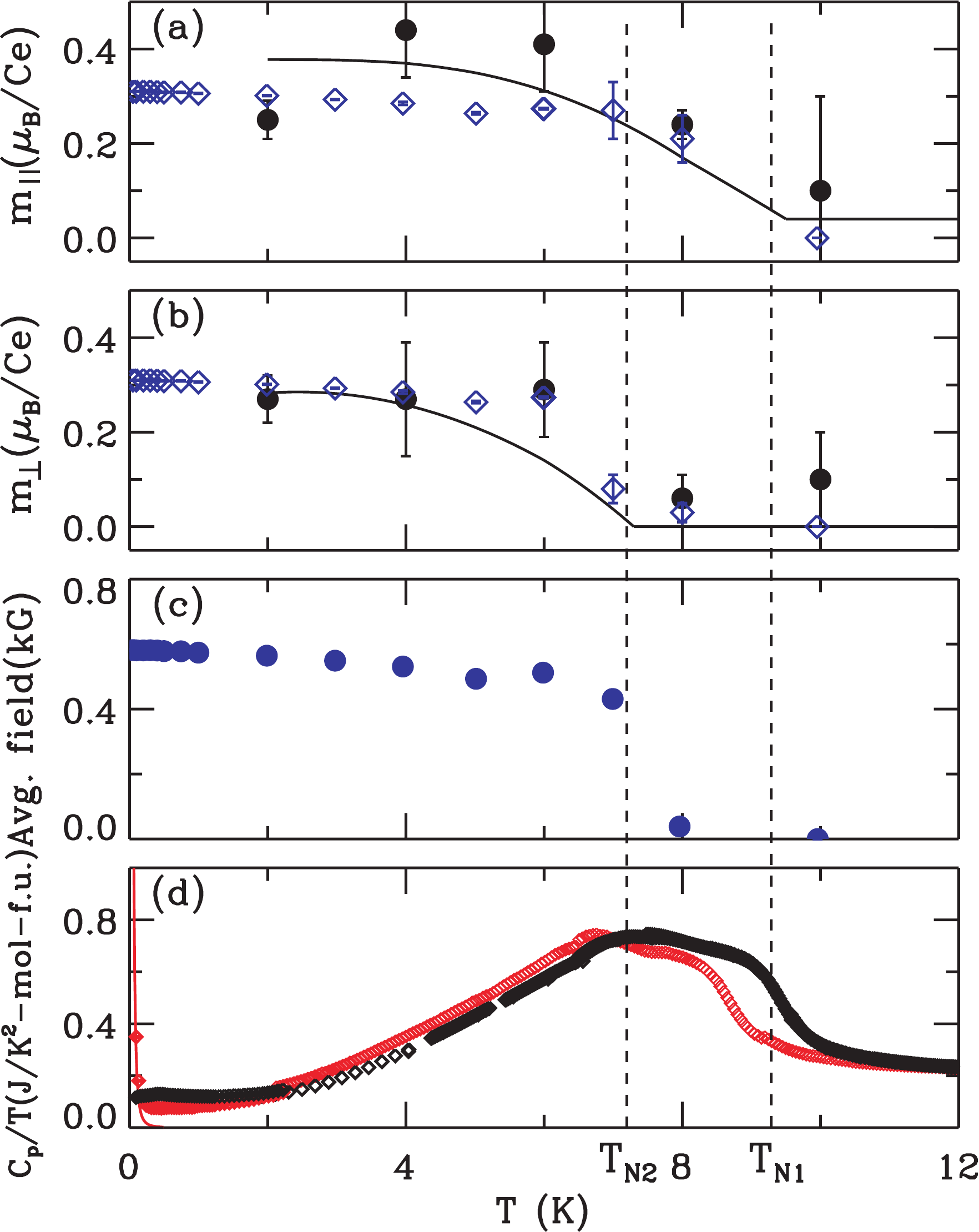}
\caption{\label{tdp}  Temperature dependence of (a) the longitudinal ($m_a$) and (b) the transverse  moments ($m_c$ for high $T$ and $m_b$ for low $T$ phase). Black dots were  extracted from Rietveld fits to neutron diffraction data. The 2 K and 8 K data points were averaged over two chopper settings. Blue diamonds were inferred from $\mu$SR fits. The solid lines are guides to the eye.   (c) Temperature dependence of the averaged static field. 
(d) Specific heat $C_p/T$ in zero field and for $\mu_0 H = 14$ T. The upturn in $C_p/T$ at 14 T is due to the nuclear spin contributions as indicated by the solid red line. }
\end{figure}

We determined the fundamental magnetic wave vector and spin polarization through neutron diffraction. Weak magnetic peaks are apparent at $T$ = 2 and 8 K after subtracting data at $T$ = 15 K (Fig. \ref{fig2} (c-d)). At $T$ = 2 K,  the difference pattern shows several resolution limited peaks. The peak with the lowest wave vector transfer  $Q \approx $ 0.77 $\rm \AA^{-1}$  can be  indexed as ${\bf Q}_m = (0.5, 0, 0)$. Magnetic neutron diffraction probes spin polarization perpendicular to wave vector transfer so this indexing implies spin  components along   \textbf{b} and/or \textbf{c}. 
Upon warming to 8 K $< T_{N1}$, the absence of  this first peak is indicative of a  longitudinal spin density wave (SDW) polarized along \textbf{a}.  The width of the intensity maxima for $T = 8$ K and $Q \approx$ 1.1 $\rm \AA^{-1}$ in Fig. \ref{fig2} (c)  exceeds the  instrumental $Q$-resolutions. The incommensurability indicated by  $\mu$SR can account for this. 
The magnetic signal at 8 K is however quite weak and since there is no energy resolution, inelastic magnetic scattering may also contribute to the broadened peaks, particularly near the  polarization suppressed ${\bf Q}_m$ peak. The diffraction data thus do not permit a unique determination of the spin structure for $T_{N2} <T<T_{N1}$.The combination of muon, specific heat, and elastic/inelastic neutron data, however, does allow an accurate determination of both structures. 

Using Kovalev notation \cite{kova,sarah}, the reducible magnetic representation associated with $\textbf{k}=(\mu 0 0)$ decomposes into three two-dimensional irreducible representations (IR): $\Gamma_{mag} =2\Gamma_1^{(2)} +\Gamma_2^{(2)}$ with 6 Basis Vectors (BVs) (Table~S2). Landau theory allows only one IR for  each of the two second order phase transitions. Below $T_{N1}$, BVs $\psi(4)$ and $\psi(6)$ of $\Gamma_1$ depict a spin structure with moments  along {\bf a}. Adding $\psi(3)$ and $\psi(5)$  allows for moments along {\bf c}.  
 Below $T_{N2}$, we can account for the diffraction pattern in Fig.  \ref{fig2} (d) by adding  $\psi(1)$ and $\psi(2)$ of $\Gamma_2$. 
The best fit corresponds to a reduced $\chi^2$ = 1.95 and a staggered moment $\langle m \rangle =$ 0.37(5) $\mu_B/\rm Ce$  that is canted by  $\varphi \approx 36(6)^\circ $ to the \textbf{a} axis (Fig. \ref{structurepd} (c)). While allowed by symmetry, the diffraction data place a limit of 0.06 $\mu_B$ on any {\bf c}-component of the staggered moment.

$\mu$SR, which probes magnetism in real space, offers an independent assessment of the proposed structures. We find a consistent description of the precession data with the muon stopping site ($\frac{1}{4},\frac{3}{4},z_{\mu}$) in Fig.\ref{structurepd} (a). The fitting analysis described below yields $z_{\mu}=0.1471(3)$ (=$z_{\rm Ce}$), close to the preferred  distance of muons  from $\rm O^{2-}$ \cite{musrO}. This location is also favored considering the electrostatic potential-energy map for CeFeAsO \cite{musrReFeAsO}. 
The observation of two muon  precession frequencies suggests two magnetically inequivalent muon sites (see Fig.\ref{structurepd} (b-c)). 
The asymmetry pattern $P_\mu^z(t)$ can be fitted to equation S1 wherein the magnetic field distribution function $\rho_i(B)$ is calculated directly from the spin structures. For the low $T$ commensurate state, $\rho_i(B)$  consists of two delta functions corresponding to the magnetic field at each of the two magnetically inequivalent (but crystallographically equivalent) muon sites. 
The best fit  is obtained with moment $m = 0.37(2) \mu_B$ and rotation angle $\varphi = 36(7)^\circ $, 
which is in excellent agreement  with the Rietveld refinement of neutron diffraction.
For the high $T$ incommensurate state, $\rho_i(B)$ is continuous: The incommensurate nature of the spin structure ensures  every muon site, though crystallographically equivalent, is magnetically unique and  contributes a distinct precession frequency. The best fit leads to an incommensurate wave vector {\bf k}=(0.44(4),0,0), $m_a=0.27(6) \mu_B$, and $m_c=0.08(3) \mu_B$. The corresponding calculated  muon asymmetry and neutron diffraction are in Fig.  \ref{fig2} (a)$\&$(c). A small component of $m_c$ implies this is a magnetic cycloid.  The corresponding lack of  inversion symmetry could have interesting consequences for electronic transport. However, since $m_c$ $\ll$ $m_a$ we retain the terminology of a longitudinal SDW. 
In summary, the  spin structures  for two ordered states  -- a longitudinal SDW (Fig. 1 (b), Fig. S4) and a commensurate coplanar structure (Fig. 1 (c)) -- account for both neutron and $\mu$SR data. 

\begin{figure}[t]
\includegraphics[width=0.8\columnwidth,clip,angle =90]{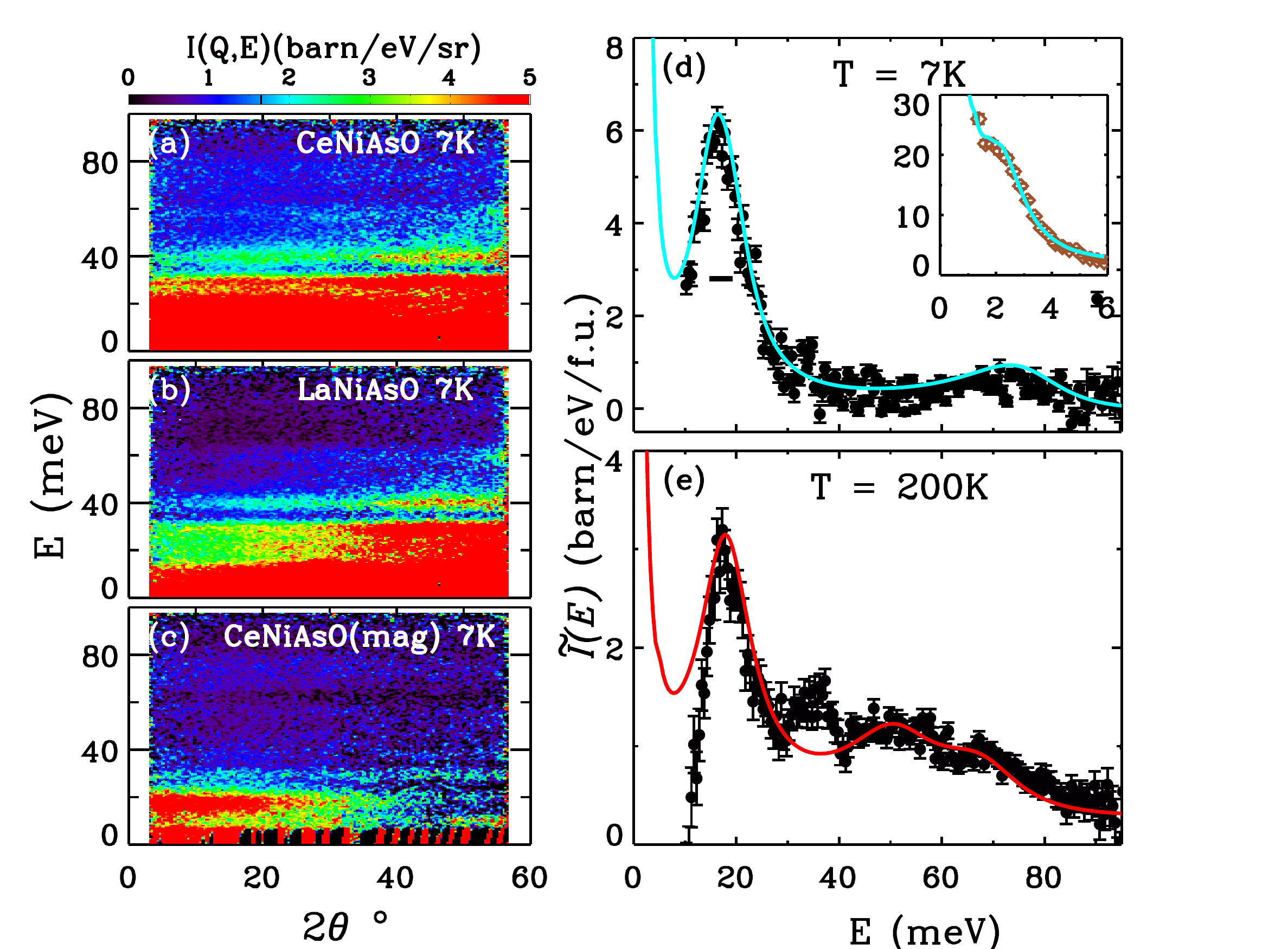}
\caption{\label{slice}  Normalized inelastic spectrum with incident energy $E_i = 100 $ meV (black dots) for (a) CeNiAsO and (b) the non-magnetic reference LaNiAsO. (c) The $T$ = 7 K difference spectrum: $\tilde{I}(Q,E)=I_{Ce}-rI_{La}$ where $r = \sigma_{\rm CeNiAsO} / \sigma_{\rm LaNiAsO}  $. (d-e) Momentum-integrated scattering 
at $T$ = 7 K and 200 K by using the method in Ref. \cite{murani1983,murani1994,osborn1993}.
The horizontal black bar indicates energy resolution. \color{black} The inset in (d) shows a   magnetic excitation at $ 2 $ meV in the ordered state with $E_i = 50 $  meV (brown dots).  The cyan and red solid lines were calculated for the crystal field model  described in the text. }
\end{figure}

For context we examine the $4f$ electron crystal field excitations through inelastic magnetic neutron scattering (Fig. 4(a-e)). At $T$ = 7 K, the intensities of modes at E $\approx$ 10 meV, 30 meV  and 40 meV rise with $Q^2$ and are observed  both for CeNiAsO and non-magnetic LaNiAsO and so must be vibrational\cite{neutron_cs}.
In the difference data $\tilde{I}(Q,E)$ and $\tilde{I}(E)$  (Fig. \ref{slice} (c-e)), we associate the two broad modes at $E_1 \approx 18(3)$ meV and $E_2 \approx$ 70(8) meV with magnetic excitations because their intensity decreases with $Q$ as the $4f$  formfactor. In the tetragonal environment of $\rm Ce^{3+}$, the $J = \frac{5}{2}$ multiplet splits into three Kramer's doublets. The two magnetic modes are correspondingly assigned to crystal-field-like excitations from the ground state (GS) to two excited doublets.  At $T$ = 200 K population of the excited state yields a broad mode at $50$~meV$\approx E_2-E_1$, which arises from excitations between the excited doublets.  Finally we observe a sharp mode at $E_{0} \approx 2 $ meV within the AFM ordered state (inset, Fig.\ref{slice} (d)). In the language of CEF theory, this is an intra-doublet transition driven to inelasticity by the molecular exchange field. As expected for a strongly correlated solid, the crystal field excitations measured for a powder sample are  broadened by damping and dispersion, leading to the half width at half maximum  (HWHM) of $\Gamma_1 = 13 ~$ meV and  $\Gamma_3 = 24 ~$ meV. 
Fitting to Lorentzian spectral functions leads to HWHM of $\Gamma_0 = 2 ~$ meV that is comparable to the Kondo temperature $T_K=15(5)~$K inferred from thermo-magnetic data \cite{Luo2012}. \color{black}

Given these broad modes, a local moment crystal field model cannot be comprehensive but it provides a useful starting point. As detailed in the SI, we carried out a global fit of a symmetry-constrained crystal field model to the normalized scattering data $\tilde{I}(E)$ at $T = $ 7 K and 200 K. After optimizing the crystal field parameters a molecular exchange field and three transition specific relaxation rates, Fig. \ref{slice}(d-e) shows a consistent description of data from two instrumental configurations and two temperatures is achieved. The model also accounts for the temperature dependent susceptibility data. 
Consistent with the easy plane (\emph{ab} plane) character of the ordered states, the GS wave function is  $|\pm\frac{1}{2}\rangle$ ($\Gamma_7$). 

As indicated in the DFT FS plot (Fig. \ref{structurepd} (d)), the ordering wave vector ${\bf k}= (0.44(4), 0, 0)$ satisfies a nesting condition. This suggests the ordered state for $T_{N2}<T<T_{N1}$  should be classified as a SDW \cite{Bao_RP2000, CeRhIn5_prl2015,Cr_RMP,Sidis,Mazin}. It is common for incommensurate (IC) magnets to undergo a longitudinal to transverse spin reorientation transition that reduces the modulation in the magnitude of the dipole moment per unit cell while sustaining the IC modulation \cite{Cr_RMP,NVO}. The situation is different for CeNiAsO, which not only develops  transverse magnetization but also becomes commensurate for $T<T_{N2}$. To arrive at the spin structure in Fig. \ref{structurepd} (c) from the commensurate version of Fig. \ref{structurepd} (b) involves counter-rotating the upper and lower AFM layers of a CeO sandwich  (Fig. \ref{structurepd} (a)) by $\varphi = 36^\circ (5)$ around {\bf c}.  While inter-layer bi-linear interactions vanish at the mean field level for $ \textbf{k} = (0.5, 0, 0)$ type order, inter-layer bi-quadratic interactions\cite{motome_2012,motome_2014}  give rise to a term in the free energy of the form $(m^2 \cos 2\varphi)^2$ that can favor $\varphi=45^o$ for a commensurate structure only. As $m$ grows upon cooling this term can be expected to induce both the IC to commensurate transition and the symmetry breaking transverse magnetization at $T_{N2}$. 

This brings us to the character of magnetism in $\rm CeNiAs_{1-x}P_xO$. Upon cooling, CeNiAsO passes from Fermi liquid to IC SDW to commensurate non-collinear order in two second order phase transitions. P doped samples   that we  examined ($\rm CeNiAs_{1-x}P_xO$ for $x>$ 0.1) all show the characteristic $\mu$SR oscillation associated with IC magnetism (Fig.\ref{fig2} (a)) down to 50 mK. This indicates the commensurate state is limited to a low  $T$, low $x$ pocket (Fig.~\ref{structurepd} (d)) and the initial instability of the strongly correlated Fermi liquid in $\rm CeNiAs_{1-x}P_xO$ is to an IC SDW. An important open question is whether the characteristic wave vector of the SDW evolves with $x$ or continues to be associated with the small FS as for $x=0$.

We thank Gerald Morris, Bassam Hitti, and Iain McKenzie for $\mu$SR beam line support, Sky Cheung and Yipeng Cai for assistance during the $\mu$SR experiment, and Allen Scheie for useful discussions on crystal field analysis. This research was funded by the US Department of Energy, office of Basic Energy Sciences, Division of Material Sciences and Engineering under grant DE-SC0019331 and by the Gordon and Betty Moore foundation through GBMF4532. The research at ORNL's Spallation Neutron Source was sponsored by the Scientific User Facilities Division, Office of Basic Energy Sciences, U.S. Department of Energy. Activities at Columbia and TRIUMF were supported by NSF DMR-1436095 (DMREF) and DMR-1610633, and the REIMEI project of Japan Atomic Energy Agency.


\bibliography{bibfile}
[48] See Supplementary Material for  the following information: the sample synthesis \cite{mcqueen2009}, details in the Rietveld refinement \cite{fullprof,zhao_nature2008}, details in the DFT calculation \cite{Kresse1, Kresse2, Kresse3}, analysis of Schottky anomaly  \cite{Kittel, Heltemes, Bredl_PRL} in the low-T specific heat, crystal field analysis and corresponding magnetic susceptibility calculation \cite{Steven, Lovesy, jsen}.

%
%
%
%
\end{document}



\title{Supplementary Material for `Incommensurate magnetism near quantum criticality in CeNiAsO'  }

\author{Shan Wu$^{1, 2*}$}

\author{W. Adam Phelan $^{1}$}

\author{L. Liu$^{3}$}

\author{J. R. Morey$^{1}$}
\author{J. A. Tutmaher$^{1}$}

\author{J. C. Neuefeind$^{4}$}
\author{Ashfia Huq$^5$}
\author{Matthew B. Stone$^{5}$}

\author{M. Feygenson$^{6}$ }

\author{David W. Tam$^{7}$}

\author{Benjamin A. Frandsen$^{8}$}

\author{Benjamin Trump$^{9}$}

\author{Cheng Wan$^{1}$}

\author{S. R. Dunsiger$^{10}$}

\author{T. M. McQueen$^{1,11}$}

\author{Y. J. Uemura$^{3}$}

\author{C. L. Broholm$^{1,11,12}$}

\date{\today}
\begin{abstract}
In support of the main text, the following supplementary information is provided: (1) synthesis details and nuclear structure refinement from neutron diffraction; (2) details about the neutron measurements, $\mu SR$ experiments, and DFT calculation; (3) details and discussion about an additional low temperature $T_3$ anomaly in specific heat; (4) crystal fields analysis of inelastic neutron scattering data.
\end{abstract}

\maketitle

\section{Synthesis method and nuclear structure determination}
\subsection{Synthesis method}
Samples of CeNiAsO were synthesized by pelletizing $\rm CeO_2$ (Alfa Aesar, 99.99 $\%$), NiAs (pre-fired), and Ce metal (Alfa Aesar, 99.8 $\%$) stoichiometrically, sealing the pellet in an evacuated-fused silica tube. The reaction ampoule was placed in a furnace and heated to 1000 $^\circ$C at a rate of 100 $^\circ$C/hr.  The temperature was held constant at 1000 $^\circ$C for 5 hours.  The ampoule was then allowed to furnace cool to room temperature. 

Phosphorous doped samples were synthesized following the procedure described in Ref. \cite{mcqueen2009}. For each sample,  $\rm CeO_2$, $\rm Ni_5P_4$ (pre-fired), Ce metal, purified and dried P, and NiAs (pre-fired) were mixed and ground thoroughly  and stoichiometrically to $\rm CeNiAs_{1-x}P_xO_{0.98}$. The pressed pellet was placed in a dry alumina crucible and sealed into an evacuated ampoule. The ampoule was heated from 750 $^\circ$C to 1050 $^\circ$C over 1 hour and held at 1050 $^\circ$C for 5 hours. After cooling  to room temperature in the furnace, the sample was re-grounded and re-pressed with excess of 2 $\%$ P to account for vaporization during pre-reaction. The sample was then pressed into a pellet again and placed in the same alumina crucible into a new ampoule. The ampoule was sealed into a larger ampoule backfilled with 1/3 atm Ar (99.99$\%$). This ampoule was placed directly into a furnace that was pre-heated to 1300 $^\circ$C and  it remained  there for 8-12 hours. After cooling on the bench top, such heating to 1300 $^\circ$C and bench-top cooling was repeated 2-4 times with intermediate grinding, until the starting materials were no longer visible in powder x-ray diffraction (details later). The entire synthesis was conducted in an argon atmosphere. The sample was exposed to air  less than 10 mins during each regrind/reheat/reseal cycle.

\subsection{Nuclear structure determination}
\begin{figure}[t]
\includegraphics[width=0.8\columnwidth,clip,angle =90]{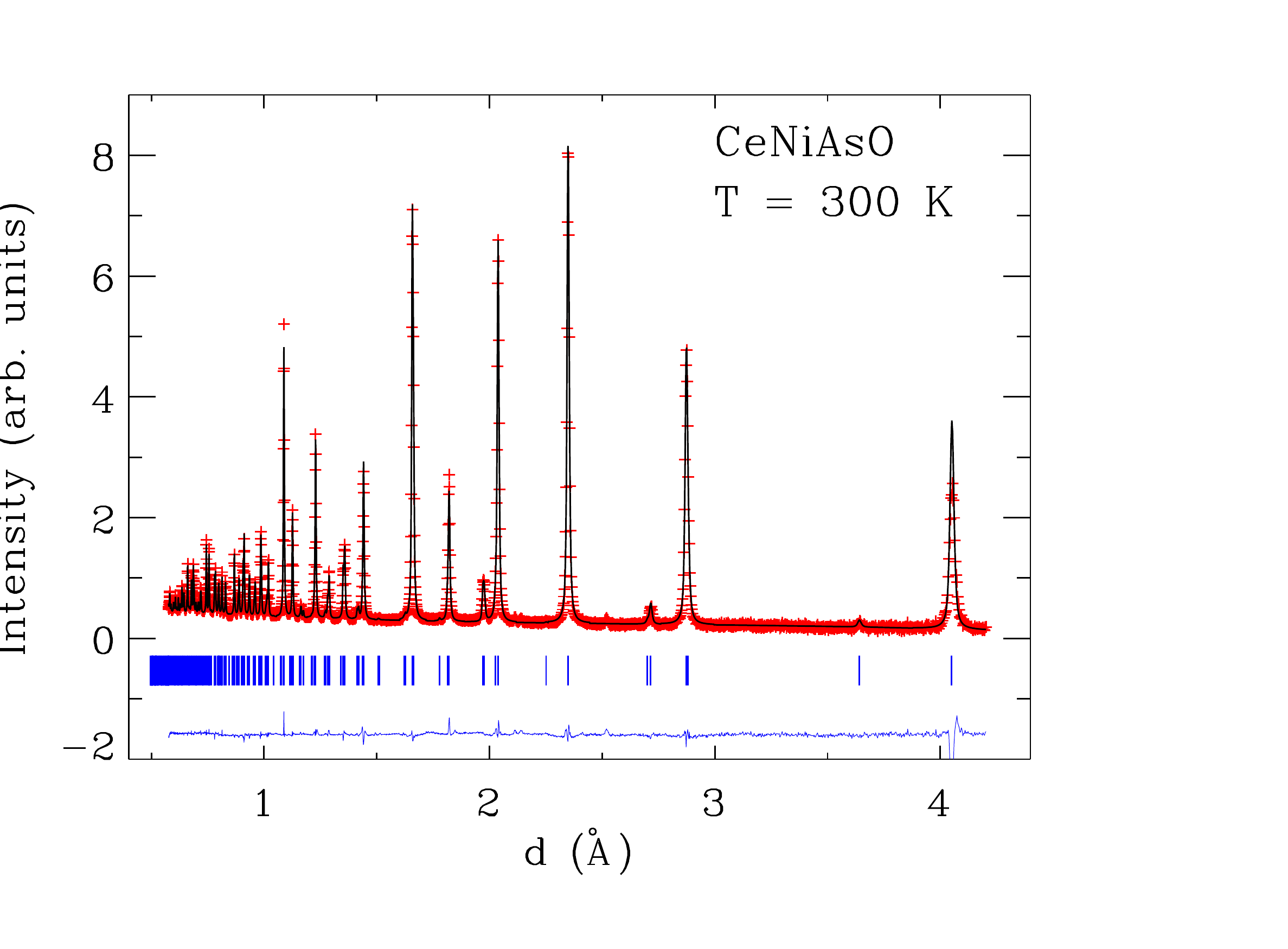}
\vspace{-1.5em}
\caption{\label{st}  Rietveld refinement of neutron diffraction that establishes the chemical structure of CeNiAsO at T = 300 K. Red dots are from POWGEN in the year of 2013 with a total proton charge of $3.3\cdot 10^{-3}$~Ah. The black curve is the Rietveld profile based on the tetragonal $\textit{P4/nmm}$ spacegroup. The blue trace is the difference between the measurements and the Rietveld profile. }
\vspace{-1.5em}
\end{figure}

The  structure of CeNiAsO was determined by Rietveld refinement of $T$ = 300 K data acquired on POWGEN (Fig. \ref{st}), using  \textit{Fullprof}\cite{fullprof}. The data were refined in the tetragonal space group \textit{P 4/nmm}. The corresponding atomic positions and lattice parameters are listed in Table \ref{st} (Bragg R-factor = 5.37).  The inferred structure is consistent with powder x-ray diffraction (XRD) refinement.  
Laboratory powder x-ray diffraction data were collected using a Bruker D8 Focus diffractometer equipped with Cu $K\alpha$ radiation and a LynxEye detector.  Phase identification and phase purity checks were conducted through Rietveld refinement. A 4 hours  XRD measurement found only $0.75\%$ by volume of $\rm CeNi_2As_2$ as the only detectable impurity phase. A pair distribution  analysis of the NOMAD data found no evidence of a local structural distortion as in CeFeAsO \cite{zhao_nature2008}. In CeNiAs$_{1-x}$P$_x$O samples with nominal values of $x$=0.2, 0.3, 0.4, and 0.5, we found $x=$ 0.19(1), 0.29(1), 0.33(1) and 0.49(1)  respectively. These values were consistent with SEM/EDX measurements.

\begin{table}
 \setlength{\tabcolsep}{8pt}
\caption{ $T = 300$~K atomic positions for CeNiAsO with space group \textit{P4/nmm}  determined by the Rietveld refinement of neutron powder diffraction data. The corresponding lattice parameters are $a=4.0621(1)$ \AA  and $c=8.1058(3) $ \AA .  }   \label{ST300}
  \begin{tabular}{ c || c| c | c | c | c  }
  \hline
Atom & Site & $x$ & $y$ & $z$ & $U_{iso}(\AA)$  \\ [0.1cm] \hline 
 Ce & $2c$ & 1/4 & 1/4 & 0.1471(3) & 0.0070(5)\\ [0.1cm] 
   Ni &  $2b$ & 3/4 & 1/4 & 1/2 & 0.0120(3)\\ [0.1cm] 
   As &  $2c$ & 1/4 & 1/4 & 0.6439(3) & 0.0081(4)\\ [0.1cm] 
  O &  $2a$ & 3/4 & 1/4 & 0 & 0.0056(4)\\[0.1 cm]\hline
  \multicolumn{6}{c}{$R_{wp} = 7.07$, $R_p=10.3$,$R_{exp}=2.06$, $\chi^2 =10.9$} \\\hline
  
    \end{tabular}
   \vspace{-1.em}
\end{table}

\section{Experimental details}
\subsection{Elastic scattering}
Neutron diffraction was carried out using the Nanoscale Ordered Materials Diffractometer (NOMAD) at the Oak Ridge National Labratories (ORNL). Data were collected with a total proton charge of 4.5 mAh at $T$ = 2 K, 8 K, and 15 K, with a 30 Hz  bandwidth (BW) chopper, admitting a wavelength band from 0.1 $\rm \AA$ to 3 $\rm \AA$ in the first frame. Temperature dependence measurements were obtained with a proton charge of  2 mAh each from 2 K to 10 K in 2 K steps, with 60 Hz pulse rate and admitting a wavelength band from 3 $\rm\AA$ to 6 $ \rm \AA$. We used a vanadium sample can and corrected for background through separate measurements of the empty instrument and an empty sample can. Diamond scans were used for calibration purposes.   Absolute unites for magnetic diffraction were obtained by scaling to nuclear Bragg intensities.

\begin{table}[t]
\setlength{\tabcolsep}{8pt}
\caption{\label{IR}The 6 basis vectors associated with magnetic structures that transform according to irreducible representations with propagation vector \textbf{k} = ($\frac{1}{2}$, 0, 0), using Kovalev notation 
as implemented in SARAh.}
\begin{tabular}{|c|c|c|c|}
\hline
IR & BV & Atom1 & Atom2 \\
& & $m_x$$m_y$$m_z$ & $m_x$$m_y$$m_z$ \\ \hline
\multirow{2}{*}{$\Gamma_2$} & $\psi(1)$ & 0 0 0 & 0 -1 0 \\ \cline{2-4}
& $\psi(2)$ & 0 1 0 & 0 0 0 \\ \hline
\multirow{4}{*}{$\Gamma_1$} & $\psi(3)$ & 0 0 1 & 0 0 0 \\ \cline{2-4}
& $\psi(4)$ & 0 0 0 & 1 0 0 \\ \cline{2-4}
& $\psi(5)$ & 0 0 0 & 0 0 -1 \\ \cline{2-4}
& $\psi(6)$ & 1 0 0 & 0 0 0 \\ \hline
\end{tabular}
\vspace{-1.em}
\end{table}

\subsection{Inelastic scattering}
Inelastic neutron scattering was performed using the SEQUOIA Fermi chopper spectrometer at SNS, ORNL. We used a multi-sample exchange system to mount CeNiAsO (m $\approx 2.31 $ g) and its non-magnetic analogy LaNiAsO (m $\approx 2.32$ g) in two thin cylindrical aluminum cans, sealed under 1 atm  of $\rm ^4He$. The sample was cooled to $T$ = 7 K using a close-cycle cryostat. To optimize intensity and resolution, we used two configurations  with different incident energy $E_i$: 1) Fermi fine chopper frequency $\nu_2 = 360 $  Hz and  $T_0 $ chopper frequency $\nu_0=$ 90 Hz for $E_i$ = 50 meV;   2) Fermi sloppy chopper frequency $\nu_1 = 240  $ Hz and  $\nu_0 =$ 60 Hz for $E_i$ = 100 meV. The corresponding Full Width at Half Maximum (FWHM) elastic energy resolutions $\Delta E$ were 1 meV and 5.7 meV respectively. 

\subsection{$\mu SR$ experiment}
 We performed zero field (ZF) $\mu SR$ with the detectors in the longitudinal configuration on the M15 beam line at Canada's national laboratory for particle and nuclear physics (TRIUMF). The samples were mounted using Apiezon N grease, covered with Alfa Aesar 0.025 mm thick silver foil, and cooled in a dilution refrigerator  to 0.04 K. 
 

The asymmetry pattern $P_\mu^z(t)$ can be fitted to the following expression:
 \begin{equation}
 \label{musr}
A_s\sum_i(\frac{2}{3}e^{-\lambda_i t}\int\rho_i(B) \cos(\gamma_\mu B t)dB+\frac{1}{3}e^{-\lambda_s t} ) +A_{bg}e^{-\lambda_{bg} t}
\end{equation}
Here $A_s$ is the total asymmetry stopped in the sample at $t$ = 0. We assume equal weight for the two magnetically distinct muon sites  (ratio = 1.0(2)). $\rho_i(B)$ is the magnetic field distribution at the muon site $i$. 
$\lambda_i$ and $\lambda_s$ are the rapid transverse and slow longitudinal rate respectively, and $\gamma_\mu/2\pi = 135.53$ MHz $\rm T^{-1}$ is the gyromagnetic ratio of the muon. The last term accounts for muons stopped beyond the sample.

\begin{figure}
\includegraphics[width=1\columnwidth,clip,angle =0]{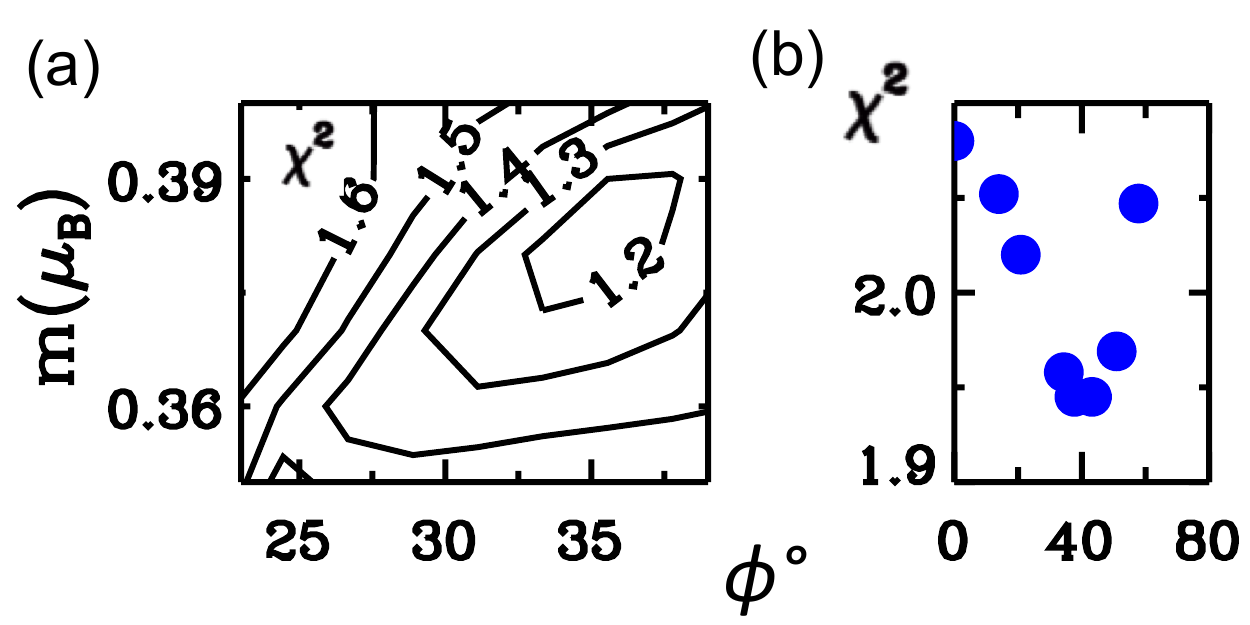}
\vspace{-1.5em}
\caption{\label{analysis} Contour plot of the reduced $\chi^2$ for $\mu$SR fits (a) and the reduced $\chi^2$ versus $\varphi$ for Rietveld fits to powder neutron diffraction data (b) below  $T_{N2}$. The rotation angle of magnetic moment $\varphi$ is defined as follows: $\rm tan\varphi = m_{\perp \textbf{k}}/m_{\parallel \textbf{k}} $.}
\vspace{-1.5em}
\end{figure}

\begin{figure}[t]
\includegraphics[width=1\columnwidth,clip,angle =0]{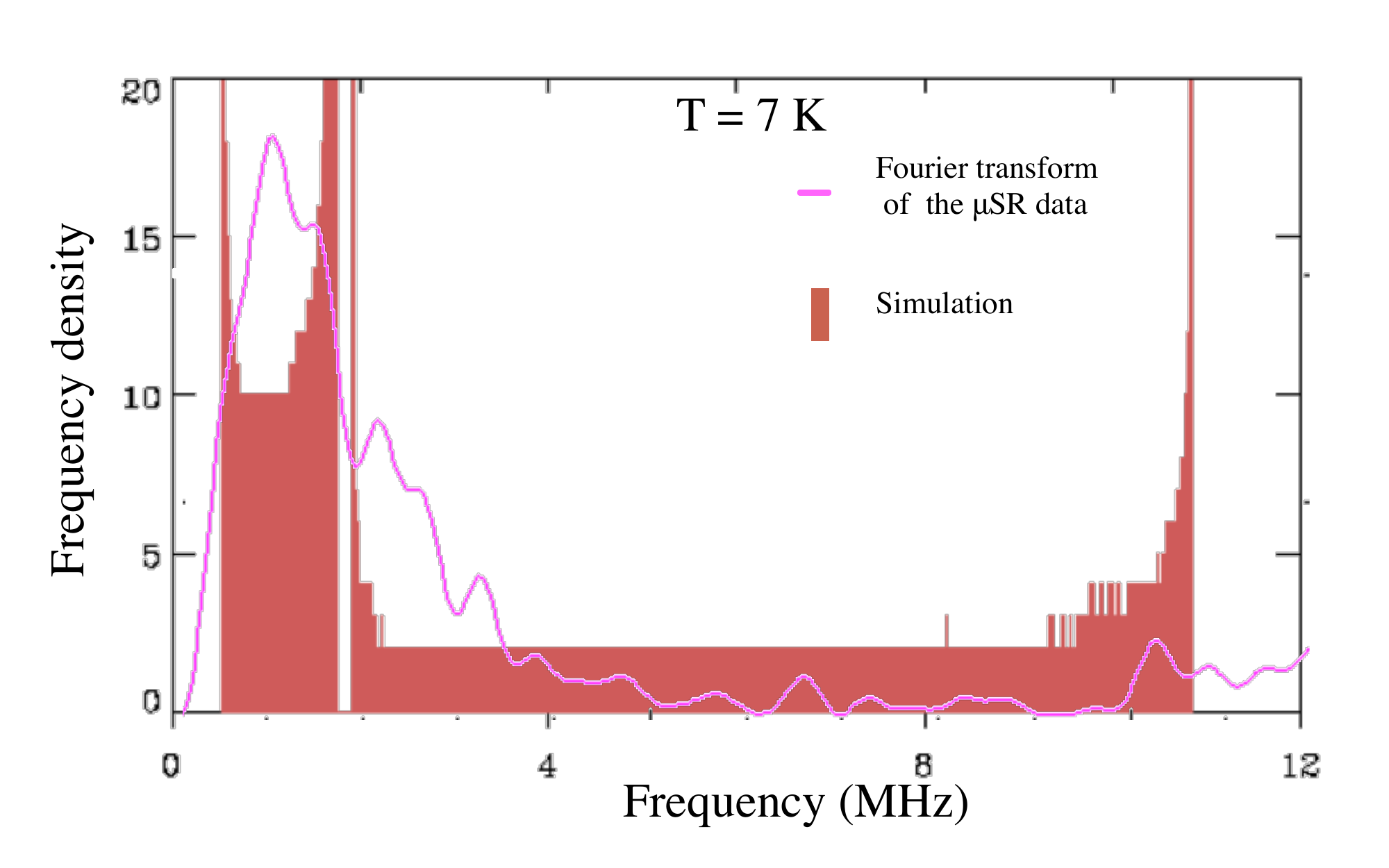}
\vspace{-1.5em}
\caption{\label{roub}   The muon ensemble precession spectrum  at $T=$7 K. The precession frequency $f$  is proportional to the internal field $B$: $f=\gamma_\mu B /2\pi $ . The solid pink line is the Fourier transformation of the experimental $\mu SR$ data. The histogram indicates the calculated spectrum for the incommensurate structure described in the main text.   }
\vspace{-1.5em}
\end{figure}

\begin{figure}[h!]
\includegraphics[width=1\columnwidth,clip,angle =0]{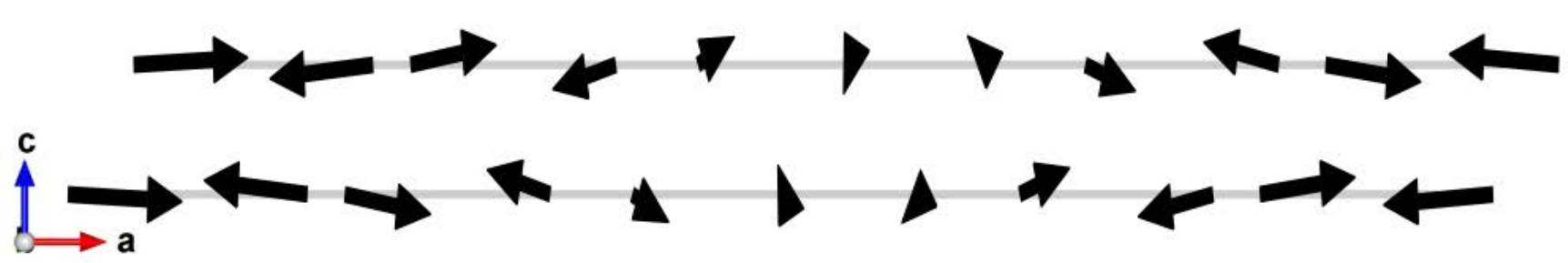}
\vspace{-1.5em}
\caption{\label{incom} Longitudinal incommensurate magnetic structure in CeNiAsO. Also shown is a small transverse component along the c-axis that is allowed by symmetry and improves the Rietveld fit to the magnetic neutron powder diffraction data.}
\vspace{-1.5em}
\end{figure}

We calculated the internal field distribution $\rho_i(B)$ on the putative muon interstitial sites and spin configurations for comparison to the $\mu SR$ data. For calculating the local magnetic field on one muon site we summed contributions from 61  crystallographic unit cells. This is sufficient to reach convergence of the internal field.  For the low-$T$ commensurate phase, plugging the value of field into Eq. S1 and comparing to the oscillation pattern, we determined  the moment $m$ and the rotation angle  $\phi$ (Fig. \ref{analysis}). The transverse and longitudinal moment $m_\parallel$ ($m\cot\phi$) and $m_\perp$ ($m\tan\phi$) was also inferred and plotted in Fig. 3 (a-b).  
For the high-$T$ phase, the frequency density  is shown in Fig. \ref{roub}. We performed an average over the phase of the incommensurate spin configuration (Fig. \ref{incom}).  From the optimal fits to  $T=7$ K $\mu SR$ data,  we determined the wave vector {\bf k},  the moment amplitude along the $a$-axis and the $c$-axis, $m_a$ and $m_c$ respectively.  Throughout the measured temperature range, the $T$-dependence of  $m_\parallel$ and $m_\perp$ obtained from the $\mu SR$ analysis are consistent with the values obtained from Rietveld refinement of neutron powder diffraction.

\subsection{DFT calculation}
Electronic structure calculations were conducted using the Vienna Ab-Initio Simulation Package (VASP), which models valence electrons using projector augmented plane waves (PAW) \cite{Kresse1, Kresse2, Kresse3}. The exchange-correlation effects were modeled using the local (spin) density approximation (L(S)DA). An 11x11x11 gamma-centered Monkhurst-Pack k-point grid was used for all calculations along with an upper energy cut-off 434 eV, and a convergence of the total energy to better than 0.01\%. The calculations proceeded in three steps. First, a structural and ionic relaxation was performed, with the ionic relaxation considered complete when the total free energy difference between steps was less than 0.1 \%. Second, a non-magnetic static calculation using this structure was performed. Finally, a magnetic static calculation initiated from the converged (non-magnetic) charge density was performed. For the magnetic calculation, an initial 0.5  $\mu_B$ magnetic moment along  $\pm {\bf c}$ was applied to the Ce sites, converging to a simplified anti-ferromagnetic structure with a moment of 0.1 $ \mu_B$ per Ce. Spin-orbit interactions and an onsite (Ce) Hubbard $U$ of 7.5 eV with an exchange ($J$) parameter of 0.68 eV were included in all calculations.


\section{Low temperature specific heat anomaly }
\begin{figure}[t]
\includegraphics[width=1\columnwidth,clip,angle =0]{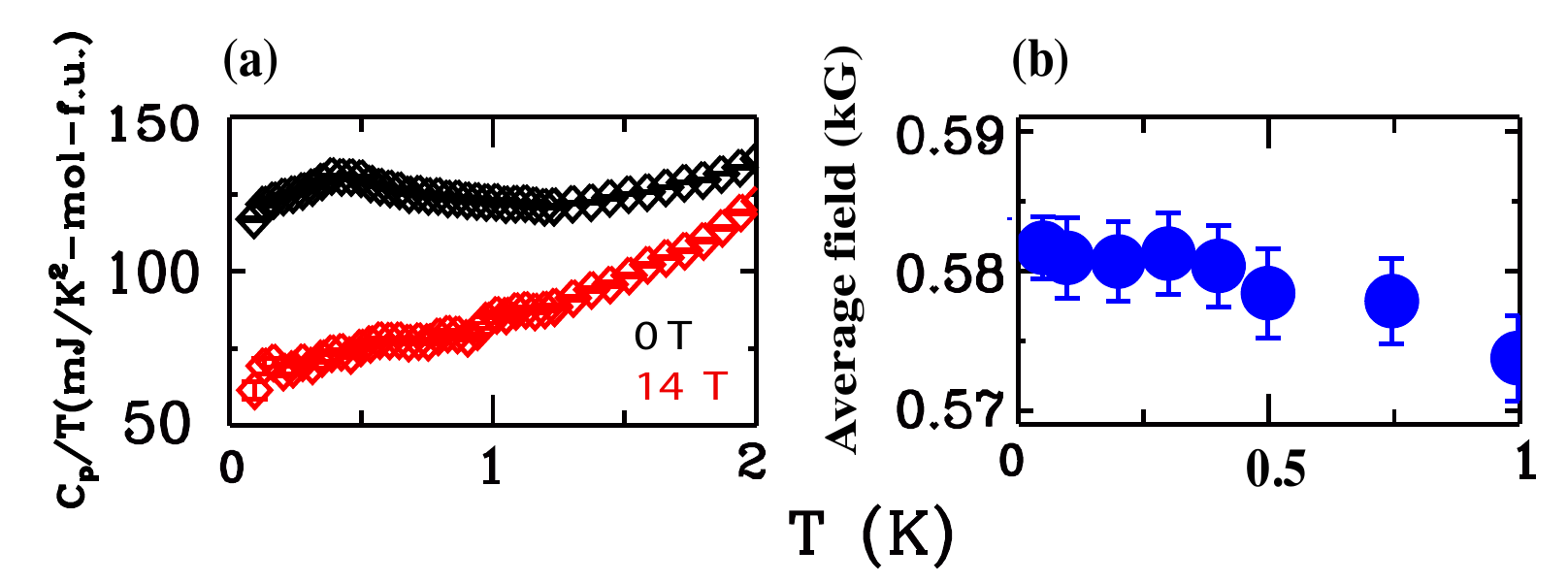}
\vspace{-1.5em}
\caption{\label{lowtcp} CeNiAsO: (a) specific heat data acquired in a PPMS plotted as $C_p/T$ for temperatures below 2 K, after subtracting the calculated hyperfine nuclear spin contributions for zero field and 15 T.  (b) Extracted average field below 1 K from muon spin rotation measurement.}
\vspace{-1.5em}
\end{figure}

Specific heat data were acquired using a quantum design Physical Properties Measurement System with a dilution refrigerator insert for measurements below 2 K. The single pulse method was employed near the ordering temperature and otherwise we used the adiabatic method. Two successive anomalies $T_{N1}$ = 9.3(3) K and $T_{N2}$ = 7.6(3) K were observed in the specific heat data as studied and discussed in the main text. The error bar for the two characteristic temperatures were estimated. A strong upturn in $C_p/T$ below 0.1 K (Fig. \ref{lowtcp} (a) ) is identified as a hyperfine enhanced $^{61}$ Ni nuclear Schottky anomaly. It can be accounted for by  the following expression\cite{Kittel,Heltemes}.
\begin{equation}
 C_N = E_N^2\beta^2R \frac{ e^{-E_N\beta} (1+e^{-E_N\beta})^4+4e^{-3E_N\beta} }{(1+e^{-E_N\beta}+e^{-2E_N\beta}+e^{-3E_N\beta})^2}
 \end{equation}
Here $E_N = -g_N\mu_N\Delta IB$ is the level splitting, where $\mu_N$ is nuclear magneton and $g_N$ is the nuclear gyromagnetic ratio. The inset in Fig. \ref{tdp} (d) shows a broad specific heat anomaly at $T_3=0.5$ K after subtracting the nuclear Schottky anomaly.  

The corresponding change in entropy, $S(\mu_0H=14$ T)-$S(\mu_0H=0$ T) $\approx$ 80 mJ/K/mol/f.u. up to 2.2 K, is $1.4\%$ of a full doublet entropy $R \ln 2$. Resistivity measurements on a polycrystalline sample  show no indication of a superconducting phase transition down to 55 mK. In addition, no clear change of oscillation patterns were observed in $\mu SR$ spectra (Fig. \ref{lowtcp} (b)). 
Possible explanations for the low $T$ specific heat anomaly  are (1) coherence effects in the Kondo lattice as for $\rm CeAl_3$ \cite{Bredl_PRL}, (2) changes in the magnetic domain wall configuration and (3) partially gapped Fermi surface with sensitivity to an applied  fields (4) an anomaly associated with a second phase in the sample.


\section{Crystal field analysis}
\begin{figure}[t]
\includegraphics[width=1\columnwidth,clip,angle =0]{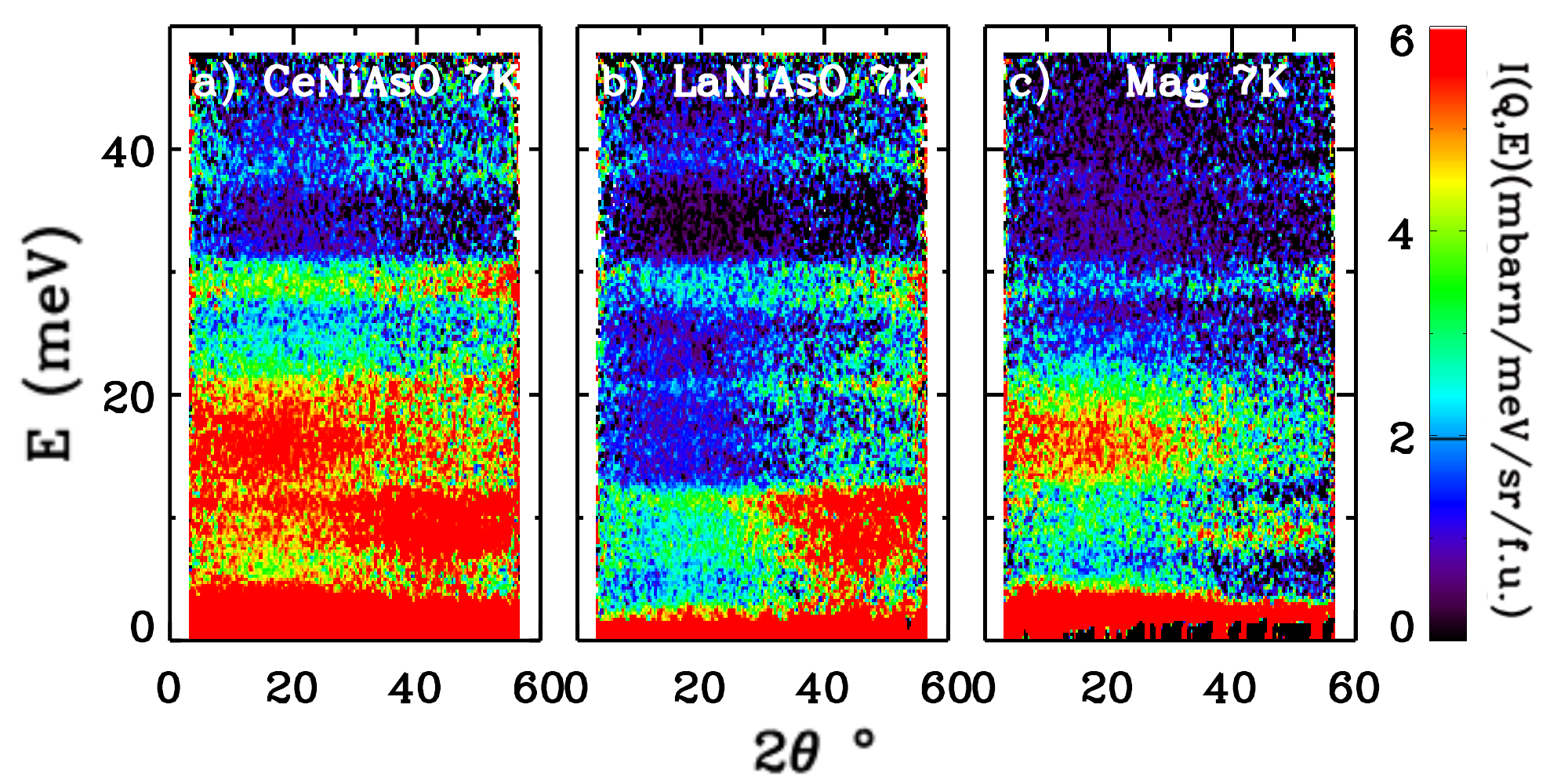}
\vspace{-1.5em}
\caption{\label{cef2}  Normalized inelastic neutron scattering cross section acquired with incident energy $E_i = 50 $ meV for (a) CeNiAsO and (b) the non-magnetic reference sample LaNiAsO. (c) The difference data, which is predominantly magnetic neutron scattering.  }
\vspace{-1.5em}
\end{figure}

The crystal field inelastic spectrum (Fig. \ref{slice} (c)-(e)) was obtained from inelastic neutron scattering data acquired on SEQUOIA after subtracting scaled data from the non-magnetic reference sample LaNiAsO and normalizing based on Bragg diffraction. In the PM phase the Hamiltonian is expressed in terms of Stevens operators [30] that are allowed by the point group symmetry and account for the observed CEF modes. In the ordered state, a mean field description of inter-site interactions adds a molecular field term. The inelastic spectrum collected with incident  energy $E_i = 50$ meV in Fig 4. (d-e) is shown in the Fig. \ref{cef2}. The $E$= 30 meV mode in  Fig. \ref{cef2}(c) is the remnant of a phonon mode that is more apparent in Fig. \ref{cef2}(a) and Fig. \ref{cef2}(b). The difference of momentum-integrated scattering $\tilde{I}(E)$  in the Fig. 4 (e) is obtained by subtracting LaNiAsO data from CeNiAsO data.\color{black} Broad CEF excitations and one sharp mode that appears in the ordered state can be reproduced by the following phenomenological model.
 The Ce atoms are located at the $2c$ crystallographic site which has $C_{4v}$ point group symmetry. In the paramagnetic state, the CEF Hamiltonian that operates on the ground state $J$-multiplet can be written as:
 
\begin{equation}
\label{eq1}
 \hat{H}_{CEF} = B_2^0\hat{O}_2^0+B_4^0\hat{O}_4^0+B_4^4\hat{O}_4^4.
\end{equation}

 Here $B_l^m$ are the  pre-factors of Stevens operators (denoted CEF parameters here) and $\hat{O}_l^m$ are the Stevens equivalent operators \cite{Steven}. In the ordered state,  the contribution of a molecular exchange field was added to the CEF Hamiltonian $\hat{H}_{CEF}$ as follows:

 \begin{equation}
 \label{eq2}
 \hat{H} = \hat{H}_{CEF}-g_j\mu_B {\bf B}_{\rm eff}\cdot{\bf J}.
 \end{equation}
 
  The cross section for unpolarized neutron scattering from a polycrystalline sample arising from transitions from state $ |i\rangle$ to $|j\rangle$ can be expressed as follows \cite{Lovesy}:
  \begin{equation}
  \label{eq3}
  \begin{aligned}
  \frac{d^2\sigma}{d\Omega dE} = & \ N\frac{k_f}{k_i}|\frac{1}{2} g_j r_0 F({\bf Q})|^2\\
 &  \ \times \frac{2}{3}\sum_{i,j,\alpha}p_i| \langle i|J_\alpha|j\rangle|^2\delta(E_i-E_j+\hbar\omega)  \\ 
  \end{aligned}
  \end{equation}
 
\begin{figure}[t]
\includegraphics[width=1\columnwidth,clip,angle =0]{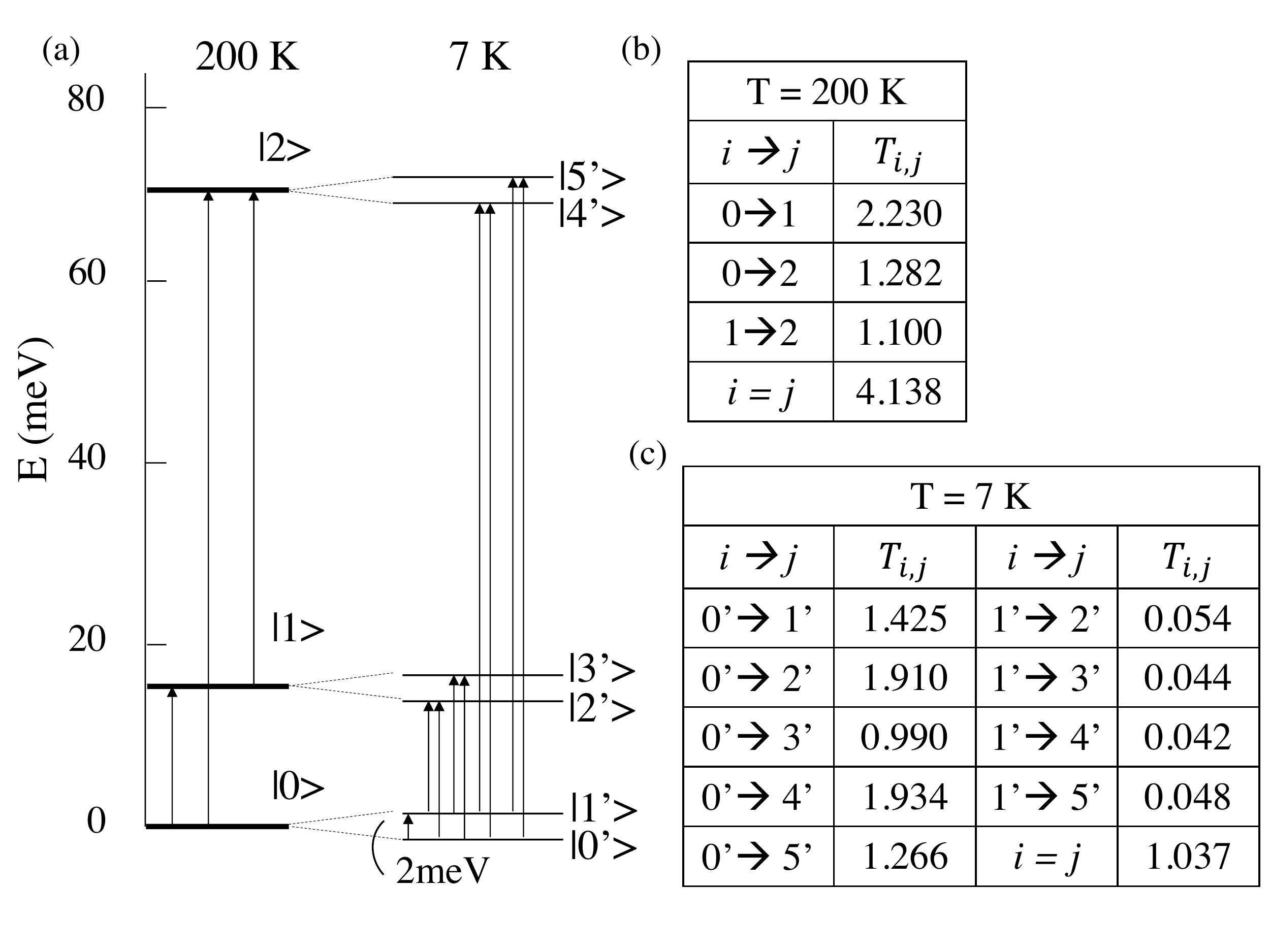}
\vspace{-1.5em}
\caption{\label{cefscheme} (a) Graphic crystal field schemes for $T =$ 200 K and 7 K. (b-c) The corresponding matrix elements $T_{i,j} = \sum_\alpha p_i |\langle i|J_\alpha|J \rangle|^2(1+e^{-E\beta}) $ from the models. The sum over all matrix elements obeys the following sum-rule $\sum_{i,j} T_{i,j} = J(J+1)$ }
\vspace{-1.5em}
\end{figure}

We replaced the Dirac delta function with a unity normalized Lorentzian function with Half Width at Half Maximum (HWHM) $\Gamma_n = 13(2), 23(3), 24(1)$ meV for the three CEF modes and $\Gamma_{0}= 2.0(2)$ meV for the sharp low energy mode. The width of three modes are much broader than the instrumental resolution even under the $E_i = 50 $ meV configuration. The dispersive bands of hybridized 4$f$ and conduction electrons may account for the broadening. The molecular field term $ {\bf B}_{\rm eff}\cdot {\bf J}$ can be further simplified as $B_{\rm eff}J_\parallel $, considering an in-plane effective field. Three CEF parameters and one molecular field parameter were obtained by fitting to the $T$-dependent magnetic neutron scattering spectra. 
The corresponding fitting parameters and wave functions are listed in Table \ref{cefpar} - \ref{wv_200K} for the PM and ordered states. The schematic crystal field scheme is displayed in Fig. \ref{cefscheme}.
The difference in  $B_2^0$ and $B_4^4$ between the two measurements is within error bars. The sign change for $B_4^0$ could result from a multipolar mean field.

\begin{table}[!h]
 \setlength{\tabcolsep}{6pt}
\caption{ The parameters obtained from fitting a crystal field model to inelastic neutron scattering data.  $B_2^0$, $B_4^0$, $B_4^4$  are coefficients of Steven operators in the crystal field Hamiltonian. $B_{\rm eff}$  is the effective molecular field within the ordered state for $T=2$~K.  }   \label{cefpar}  
  \begin{tabular}{ c || c| c | c | c}
  \hline
T (K) & $B_2^0$ (meV) & $B_4^0$ (meV)  & $B_4^4$ (meV) & $B_{\rm eff}$ (T)   \\ [0.1cm] \hline 
 7 & 3.1(1) & 0.07(1) & 1.1(1) & 1.7(2) \\ [0.1cm] 
200 & 2.3(1) & -0.09(4) & 0.9(1) & -  \\\hline
    \end{tabular}
\end{table}

\begin{table}[!h]
\caption{Wave functions considering the effective molecular field at $T$ = 7 K. The three doublets are divided into 6 non-degenerate singlets. The $\pm$ sign prior to the coefficient indicates summation over each component of angular moment.}
 \begin{tabular}{ c || c}
\hline
\label{wv_7K}
  $\psi_5$     & $\mp 0.641|\pm\frac{5}{2}\rangle \pm 0.298|\pm \frac{3}{2}\rangle \mp 0.009|\pm \frac{1}{2}\rangle$             \\  [0.1 cm]  \hline                                      
  $\psi_4$     & $- 0.648|\pm\frac{5}{2}\rangle - 0.281|\pm\frac{3}{2}\rangle + 0.008|\pm\frac{1}{2}\rangle$             \\  [0.1 cm]  \hline
   $\psi_3$     & $ \pm 0.035|\pm\frac{5}{2}\rangle \pm 0.052|\pm\frac{3}{2}\rangle \mp 0.704|\pm\frac{1}{2}\rangle$             \\  [0.1 cm]  \hline    
   $\psi_2$     & $ -0.043|\pm\frac{5}{2}\rangle + 0.079|\pm\frac{3}{2}\rangle - 0.701|\pm\frac{1}{2}\rangle$             \\  [0.1 cm]  \hline
   $\psi_1$     & $ -0.278|\pm\frac{5}{2}\rangle + 0.643|\pm\frac{3}{2}\rangle + 0.091|\pm\frac{1}{2}\rangle$             \\ [0.1 cm]   \hline
   $\psi_0$     & $\pm 0.296|\pm\frac{5}{2}\rangle \pm 0.639|\pm\frac{3}{2}\rangle \pm 0.062|\pm\frac{1}{2}\rangle$             \\ [0.1 cm]   \hline
     \end{tabular}
\end{table}

\begin{table}[t]
  \setlength{\tabcolsep}{5pt}
        \caption{Wave functions for three doublets associated with Ce$^{3+}$ in CeNiAsO inferred from inelastic neutron at  200 K.  }
\begin{tabular}{ |c|  c|   }
\hline 
Energy (meV) & Doublet wave function \\
\hline
$70(8)$        & $0.7518|\pm\frac{5}{2}\rangle+0.6593|\mp\frac{3}{2}\rangle$             \\       
\hline                                             
$18(3)$		&  $0.6593|\pm\frac{5}{2}\rangle-0.7518|\mp\frac{3}{2}\rangle$					             \\
\hline
$0$  	 &  $|\pm \frac{1}{2}\rangle$	          \\ 
\hline
    \end{tabular}

    \label{wv_200K}
    \renewcommand{\arraystretch}{1.2}
\end{table}

\begin{figure}[t]
\includegraphics[width=1\columnwidth,clip,angle =0]{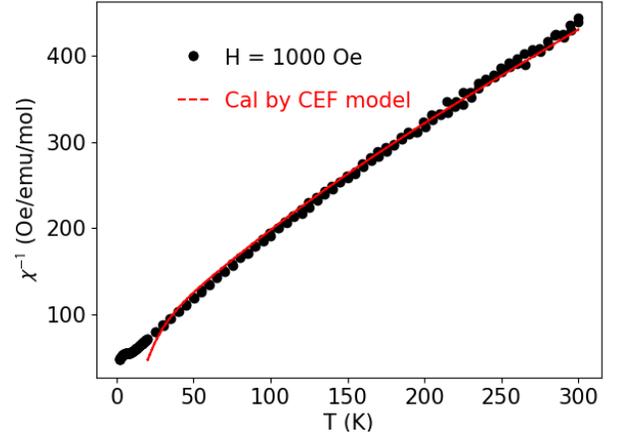}
\vspace{-1.5em}
\caption{\label{suscep} Magnetic susceptibility of polycrystalline CeNiAsO as inferred from magnetization measurements in an applied field H  = 1000 Oe. The red curve shows the calculated temperature dependent magnetic susceptibility based on the CEF model Eq. \ref{eq4} and a small diamagnetic component that was adjusted for best agreement with the data. The corresponding coefficients of the Steven operators are $B_2^0  = 3.9(1.3)$, $B_4^0 = -0.05(3)$, $B_4^4 = 0.7(1)$. }
\vspace{-1.5em}
\end{figure}

To verify the  scheme of crystal field, we also measured the magnetic susceptibility of a polycrystalline sample. The data was collected with external field H = 1000 Oe as shown in Fig. \ref{suscep}. The magnetic susceptibility $\chi^{\alpha}$ ($\alpha = x,y,z$) can be calculated as follows \cite{jsen}:
\begin{equation}
\label{eq4}
\begin{aligned}
\chi^{(\alpha)} = & N_A(\mu_B g_j)^2  \bigg(\sum_{i = j} \frac{|\langle i | \hat{J}^{(\alpha)} | j \rangle |^2}{k_B T} n_i \\
& +  \sum_{ i \neq j} \frac{|\langle i |\hat{J}^{(\alpha)}| j \rangle|^2}{E_i-E_j}  (n_j - n_i)\bigg)
\end{aligned}
\end{equation}

Here $N_A$ is Avogadro's number and $\mu_B$ is the Bohr magneton. $|i\rangle$ is the eigenstate from the crystal field analysis corresponding to the eigenvalue $E_i$. $\hat{J}^{\alpha}$ are angular momentum operators associated with the three cartesian coordinates and $n_i, n_j$ are the thermal population factors. The theoretical powder averaged susceptibility is calculated as $\chi = (2\chi^{(x)}+ \chi^{(z)})/3$. The inverse susceptibility  can then be fitted to $1/(\chi + \chi_0)$ where $\chi_0$ is a temperature independent diamagnetic term 
The corresponding fitting is shown as the red curve with CEF parameters listed in the caption of Fig. \ref{suscep} without any additional parameters adjustable parameters except for $\chi_0$ = -3.3e-3 emu/Oe/mol. This consistency with susceptibility data further reinforces our conclusion that $|\pm \frac{1}{2} \rangle$ is the Kramers doublet ground state.

We can  extract the inelastic spectral weight  for each CEF mode $E_n$ at low temperature  $T$ = 7 K as following:
 \begin{equation}
 \delta m_n^2 = 6\mu_B^2 \frac{ \int\int Q^2 \tilde{I_n}(E)(1+e^{-E\beta})/|r_0F( Q)|^2 dQdE}{\int Q^2 dQ}
 \end{equation}
$\tilde{I_n}(E)$ is momentum-integrated scattering defined in the main text. The sum of the observed inelastic spectral weight $\delta m_{n}^2$ and static moment $\langle m\rangle^2$ (= $0.37^2$ $\mu_B^2$)  spin correlation yields a total spectral weight of $m_{tot}^2 = 6.0(4)$ $\mu_B^2$ per Ce, consistent with the expected value $g_j^2 J(J+1) = 6.47$ $\mu_B^2$ with $g_j = \frac{6}{7}$.
The magnitude of $\langle \mu_x \rangle $ $=  \langle \psi_0 | g_j J_x|\psi_0\rangle$ with the ground state singlet listed in Table \ref{wv_7K} is calculated as 0.9 $\mu_B$, while the magnitude of $\langle \mu_z \rangle $ ($= \langle \psi_0 | g_j J_z |\psi_0\rangle$ ) $\sim$ 0.  This is consistent with the proposed in-plane spin structure. 

\bibliography{bibfile}